\documentclass[%
 aip,
 sd,%
 amsmath,amssymb,
 reprint,%
]{revtex4-1}

\usepackage{graphicx}
\usepackage{dcolumn}
\usepackage{bm}
\usepackage{mathtools}
\usepackage{graphicx}

\begin{document}

\preprint{AIP/123-QED}

\title{Hollow density formation in magnetically expanding helicon plasma}
\author{Sonu Yadav}
\email{syadav@ipr.res.in}
\author{Soumen Ghosh}
\author{Sayak Bose}
 \author{K. K. Barada}
 \affiliation{Institute for Plasma Research, HBNI, Bhat, Gandhinagar 382 428, India}

 \author{R. Pal}%

\affiliation{ 
Saha Institute of Nuclear Physics, I/AF Bidhannagar, Kolkata 700 064, India
}%

\author{P. K. Chattopadhyay}
\email{pkchatto@ipr.res.in}
\affiliation{Institute for Plasma Research, HBNI, Bhat, Gandhinagar 382 428, India}



\date{\today}

\begin{abstract}
Measurement of radial density profile in both the source and expansion chambers of a helicon plasma device 
have revealed that  it is always centrally peaked in the source chamber, 
whereas in the expansion chamber near the diverging magnetic field it becomes hollow above
a critical value of the magnetic field. This value corresponds to that above which both electrons and 
ions become magnetized. The temperature profile is always peaked off- axis and tail electrons are 
found at the peak location in both the source and expansion chambers. Rotation of the tail
electrons in the azimuthal direction in the expansion chamber due to gradient-B drift produces more 
ionization off-axis and creates a hollow density profile; however, if the ions are not magnetized, 
the additional ionization does not cause hollowness.

%
\end{abstract}

\maketitle

\maketitle
\section{\label{sec:level}Introduction:}
Topic of plasma expansion along diverging magnetic field is of interest 
in a number of active research fields in astrophysical \cite{ref_1,ref_2} and laboratory plasmas \cite{ref_prop2,ref_4,ref_9}. 
Plasma flow along diverging magnetic field leading to magnetic reconnection is also a well-known astrophysical
event\cite{ref_3,ref_rec}. Particularly of interest is the, helicon plasma source with expanding plasma geometry and diverging
magnetic field due to its potential application as plasma thruster \cite{ref_prop1,ref_prop2,ref_prop3} for space vehicles. 
The efficiency of thrust generation in such device critically depends on the radial
profiles of plasma density and temperature, and of electric and magnetic fields\cite{ref_azimu_Jd_1}. 
These profiles, except that of the magnetic field, have been studied experimentally \cite{ref_7,ref_8,ref_9}
and found to vary substantially on the generation of diamagnetic current, 
Hall current and additional ionization in the expanding diffusion chamber. 
In the magnetic nozzle geometry one of the interesting aspect observed\cite{ref_4,ref_5,ref_17} 
and supported by particle-in-cell simulation \cite{ref_18} is that a hollow density profile is 
generated in the expansion chamber. This is of serious concern as such type of structure
within magnetically expanding plasma causes reduction in the total thrust \cite{ref_7,ref_8} and 
is therefore necessary to avoid/remove  the hollowness of the density profile at the nozzle throat 
to increase the thrust efficiency. However, understanding of the resulting conical density
profile in the expanding plasma is still not comprehensive and demands an exploration of the 
individual roles played by the Hall current, diamagnetic current and source of ionization.

In the helicon plasma source with expanding plasma geometry and diverging magnetic field an off-axis
energetic electron component has been observed experimentally \cite{ref_4,ref_5}. These electrons, 
created by the skin heating \cite{ref_15,ref_16} in the source region near the location of rf antenna, 
are transported from the source to the expansion chamber via the last diverging magnetic 
field lines emerging from the open exit of the source chamber. 
These are speculated to have a  role in the formation of the hollow density 
profile \cite{ref_4,ref_5} by off-axis ionization. However, since the ionization length 
of these fast electrons is substantially larger than the system dimensions, 
the proposed explanation needs further investigation. The formation of the hollow structure is 
also explained in an alternative manner by the radial transport of plasma induced by radial electric 
field generated because of magnetized electrons and unmagnetized ions in the expansion chamber. 
In this mechanism an azimuthal current is proposed to be driven by the $\bold E\times \bold B$ drift (Hall current) 
which in presence of the axial magnetic field causes radial plasma transport. 
This explanation has been supported in one experiment \cite{ref_17} and by particle-in-cell simulation \cite{ref_18}. 
On the other hand, in a previous experiment \cite{ref_19} with our helicon plasma device \cite{ref_20,ref_21}
the hollow density structure is observed in the expansion chamber; however, in contrast to the above 
explanation, it is seen that the hollow profile is created even in the absence of the radial electric field. 
In this experiment \cite{ref_19} the distinct roles of the magnetic and geometric expansions were studied and found that 
it is the magnetic expansion that plays the dominant role in the hollow density formation in 
presence of the energetic tail electrons.  
In the magnetic expansion region these electrons rotate fast in azimuthal direction due to the gradient-B drift.
As a consequence their confinement is sufficiently enhanced to allow impact ionization of the neutrals 
that causes the hollow density structure.

After the crucial role of the gradient-B in hollow density formation is 
established in the previous experiment \cite{ref_19}, it would be interesting to know whether the 
absolute strength of the magnetic field has any function to play. In the present experiment 
the magnetic field is varied, which effectively changes the ion Larmor radii in both the 
source and the expansion chambers, and examine the effect. We observe that the center-peaked 
radial plasma density profile in magnetically expanding plasma transforms to the hollow profile 
when the external magnetic field strength is sufficiently increased by keeping other experimental
operating parameters same. Transition occurs when the ions becomes magnetized in the expansion 
chamber, i.e. the ion Larmor radius started becoming smaller than the chamber radius. 
So, effectively, by changing the magnetic field, the plasma density profile can be either center peaked,
or flat or hollow.  On the other hand, the density in the source chamber remains always center peaked, 
irrespective of the magnetic field of our experiment.  Our observations may lead to an effective control 
of plasma density profile with external magnetic field which may help to improve the understanding and efficiency 
of the helicon source based thrusters. 

In the next section, the experimental device and the set up for the 
experiment are described along with the diagnostics. Section III presents the experimental results 
obtained for the electron parameters in both source and expansion chambers. The results are discuss
in Section IV and finally summary and conclusion is given in section V.

\section{\label{sec:level} Experimental Setup:}
\begin{figure}
\centering
\includegraphics[width = \linewidth]{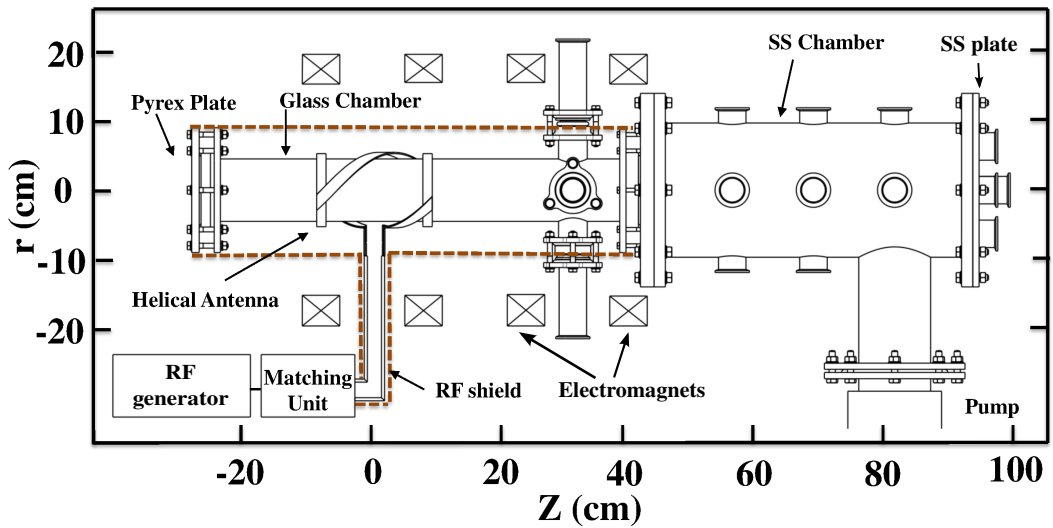}
\caption{Schematic of helicon plasma experimental setup.} \label{fig:1}
\end{figure}

\begin{figure}
\centering
\includegraphics[width = \linewidth]{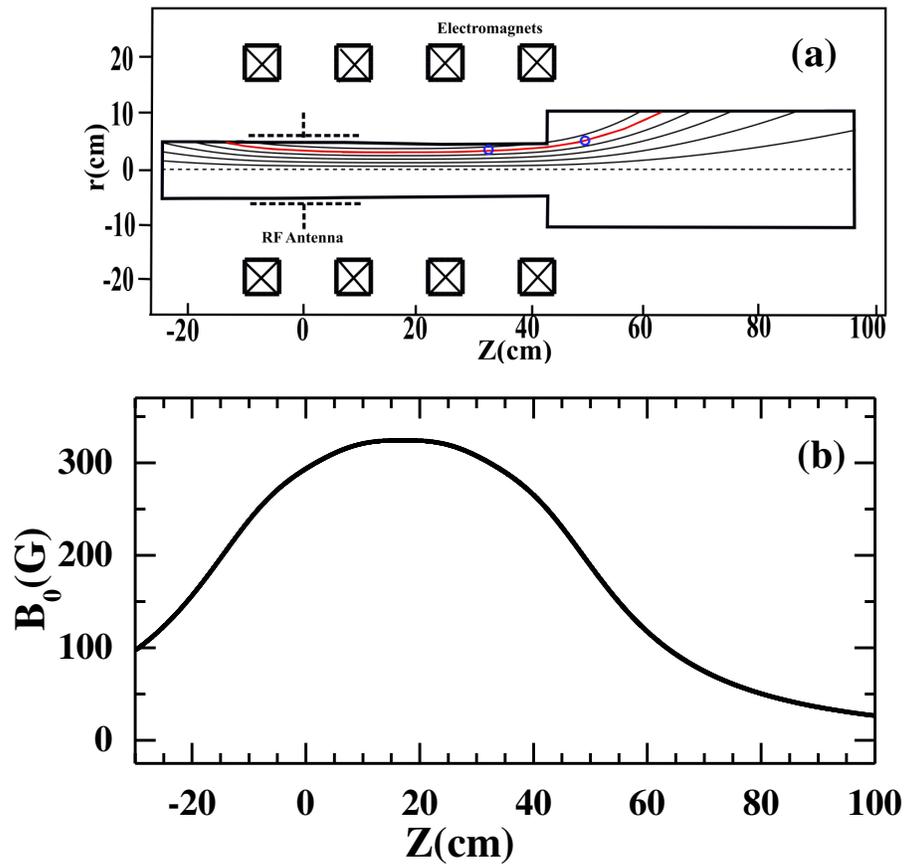}
\caption{(a) Simulated magnetic field lines and (b) field strength at 174A DC current.} \label{fig:2}
\end{figure}

The present experiment is carried out in a linear device based on helicon plasma source, shown schematically 
in Fig. \ref{fig:1}. The details is 
described in Ref. 19 and 20.
The vacuum system consists of two cylindrical chambers.
A source chamber of 9.5 cm inner diameter and 70 cm long made out of 
borosilicate glass and closed at one end with an insulating (pyrex) plate. 
The other end is connected to a 50 cm long stainless steel expansion chamber of 20 cm inner diameter, 
which is closed at the far end by a grounded SS plate.
The whole system is evacuated to a base pressure of $1\times10^{-6}$ mbar using a diffusion pump, 
connected to the expansion chamber. Argon is used as the working gas in the pressure range of
$0.7 - 3\times10^{-3}$ mbar.
An 18 cm long right helicon antenna, placed around the source chamber and 
energized by a $13.56 $ MHz radio frequency power generator through L-type impedance matching network,
produces the plasma. 
The reflected power is kept less than 2$\%$ for all the experiments.
The location of antenna center is define as $z = 0$ and all other axial locations are
with reference to the antenna center as shown in Fig. \ref{fig:1}.
Four forced water-cooled electromagnet coils, as shown in Fig. \ref{fig:1} are used 
to generate an axial magnetic field ($B_0$). 
They produce a diverging magnetic field into expansion chamber as shown in Fig. \ref{fig:2}. 
In the present experiment the electromagnet coil current (direct current), $I_B$ is varied up to 174A, 
yielding a maximum magnetic field strength of 325G at $z = 15$ cm and 290G at antenna center($z = 0$ cm). 
The simulated magnetic field lines using Poission Superfish 
software and the magnetic field strength for $I_B$ = 174A are shown 
in Fig. \ref{fig:2}a and  \ref{fig:2}b, respectively.

\begin{figure}
\centering
\includegraphics[width = \linewidth]{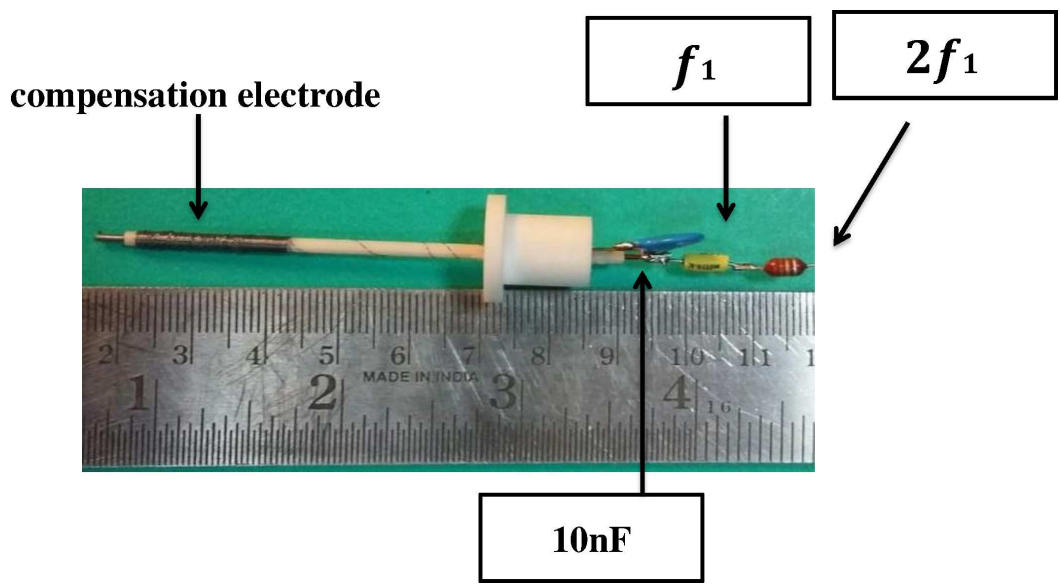}
\caption{Physical dimensions of probe tip, auxiliary electrode and self-resonating
chokes.} \label{fig:3}
\end{figure}

In order to understand the physics behind the formation of hollow density profile in the 
expansion chamber, the radial behavior of electron density, $n_e$ and temperature, $T_e$ are 
determined using rf compensated Langmuir probe \cite{ref_22}. 
The probe tip is made up of cylindrical tungsten wire of 1 mm diameter and 4 mm length. 
For rf compensation, another floating electrode is placed near it to sample the local plasma fluctuations.
The compensating electrode of 2.5 mm diameter and 20 mm length is made
of several closed winding of 0.125 mm tungsten wire over the ceramic holder of the cylindrical probe tip.  
This compensating electrode feeds the rf plasma fluctuations to 
the probe tip through a 10nF capacitor. Two tiny self-resonating chokes (at the working frequency 
$f_1$ = 13.56 MHz  
and its second harmonic 2$f_1$ = 27.12 MHz) and the compensated electrode are placed as close 
as possible to the probe tip to minimize the 
stray capacitance at rf frequencies. These self-resonating chokes are used to block any rf current entering 
the current measurement circuit.
The dimension of the compensated electrode is chosen such that it produce minimum disturbance to the local
plasma condition as well as works in a moderate plasma density of $\sim$ $10^{16}$ $m^{-3}$
in the limit of tiny self-resonating rf chokes. 
Fig. \ref{fig:3} shows the physical dimensions of the rf compensated probe.

Two different rf compensated Langmuir probes of same collection areas are used to measure profiles
perpendicular to the external magnetic field in the present experiments. 
The L-shaped rf compensation Langmuir probe is used in the expansion chamber; by rotation and translation of  
the probe it has been possible to position it in radial and axial direction, respectively. 
\begin{figure}
\centering
\includegraphics[width= 7cm]{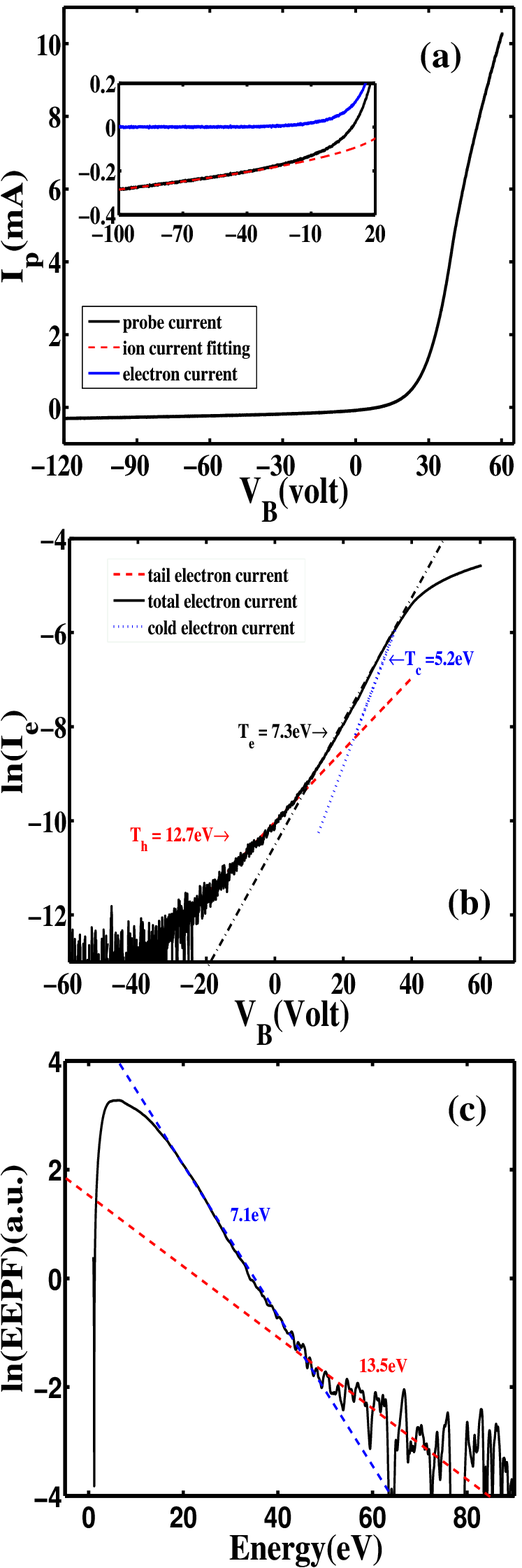}
\caption{Langmuir probe trace in Expansion chamber, at $200W, 130A, (r,z) = (5,50)cm$, 
Argon fill pressure $ = 1\times10^{-3}mbar$. (a) I-V characteristic with ion contribution (red dash line in inset), 
(b) semi-logarithmic plot of electron current and linear fit of hot (red dashed line) and cold (blue dotted line) 
electron population (c) Measured EEPF for the same condition.}
\label{fig:4}
\end{figure}

Current-voltage characteristic of the Langmuir probe is collected by sweeping the bias at the probe 
from -120V to +70V at a frequency of 2.2Hz. The data is acquired using 14 bit data acquisition system 
at a sampling rate of 100kHz with record length of 50k samples. 
Fig. \ref{fig:4}a shows a typical I-V trace of the rf compensation Langmuir probe taken in the expansion chamber
at Argon fill pressure $1\times10^{-3}$ mbar, rf power 200W and  coil current 130A. It is obtained at the 
off-axis location of $(r,z) = (5,50)$ cm in the expansion chamber. The electron temperature is 
estimated from the linear fit to natural log of exponential region of electron current after 
subtracting off the square fitted ion current (Fig. \ref{fig:4}b). 
Two separate electron populations, bulk  and high energy tail, are identified by the two straight 
line regions of semi-logarithmic plot Fig. \ref{fig:4}b. The measurement of the electron energy probability 
function (EEPF) using analog differentiator  also shows two separate electron populations at off-axis 
location in expansion chamber (Fig. \ref{fig:4}c). The temperature of the high energy tail electrons ($T_h$) 
is determined by fitting a straight line in the linear region of the semi-logarithmic plot for 
biases much more negative than plasma potential. 
This fitted line is extrapolated to plasma potential
(red dash line in Fig. \ref{fig:4}b) and subtracted from the total electron current to obtain the 
bulk electrons current.
Fitting the remaining trace to another straight line yields the bulk electron 
temperature ($T_c$) as shown by blue dotted line in Fig. \ref{fig:4}b.
The bulk and high energy tail electron densities are found by the respective electron currents at the 
plasma potential. 
The plasma potential is measured at zero crossing of analog double derivative of the I-V curve using analog 
differentiator. The fraction of electrons in the high energy tail $(\alpha)$ is equal to the current at 
the plasma 
potential due to the tail divided by the electron saturation current \cite{ref_probe_ana1,ref_probe_ana2}. 
Based on the kinetic definition of temperature, 
\begin{equation}
T_e = \frac{1}{3} m_e \int_{-\infty}^{\infty} v^2 f(v)dv
 \end{equation}
an effective electron temperature can be derived by assuming the electron energy distribution 
to be a bi-Maxwellian  
$f(v) = (1-\alpha)f_c(v)+\alpha f_h(v)$, where $f_c(v)$ is the bulk electron Maxwellian distribution and $f_h(v)$ 
is another Maxwellian distribution function due to enhanced tail of the EEDF.
\begin{equation}
 T_{e,eff} = (1-\alpha) T_c + \alpha T_h
\end{equation}
From Fig. \ref{fig:4}, we get $T_c$ = 5.2 eV, $T_h$ = 12.7 eV and $\alpha$ = 0.21, the resulting
effective electron temperature is $T_{e,eff}\simeq$7 eV. 
In our experiments $T_{e,eff}$ is nearly matches (within
10$\%$ error) to electron temperature ($T_e$) obtained without the subtraction of high energy 
tail electrons contribution
as shown by black dotted dash line in Fig. \ref{fig:4}b.
In our experiments the ratio of probe radius to Debye length also known as the Debye number 
varies within the range 3 to 10.  Laframboise theory \cite{ref_Lafram} is used to determine the ion density 
from the ion current as it is valid for the above mentioned values of Debye number \cite{ref_sayak}. 
The algorithm adopted for determination of ion density from the ion current using the 
Laframboise theory is described in Ref. 25.  
\section{\label {sec:level} Experimental Results:}
The radial behavior of electron density, $n_e$ and temperature, $T_e$, determined using the rf compensated 
Langmuir probe are obtained at two different axial locations, namely z = 31 cm and 50 cm. 
These locations are chosen such that one corresponds to before ($z_{before}$ = 31 cm), and the other 
after ($z_{after}$ = 50 cm) the magnetic and geometric expansions (Fig. \ref{fig:1}). For an electromagnet coil 
current of $I_B$ = 174A, the strength of the magnetic field at these two axial locations are 300G and 190G, 
respectively.  

\begin{figure}
\centering
\includegraphics[width= 7cm]{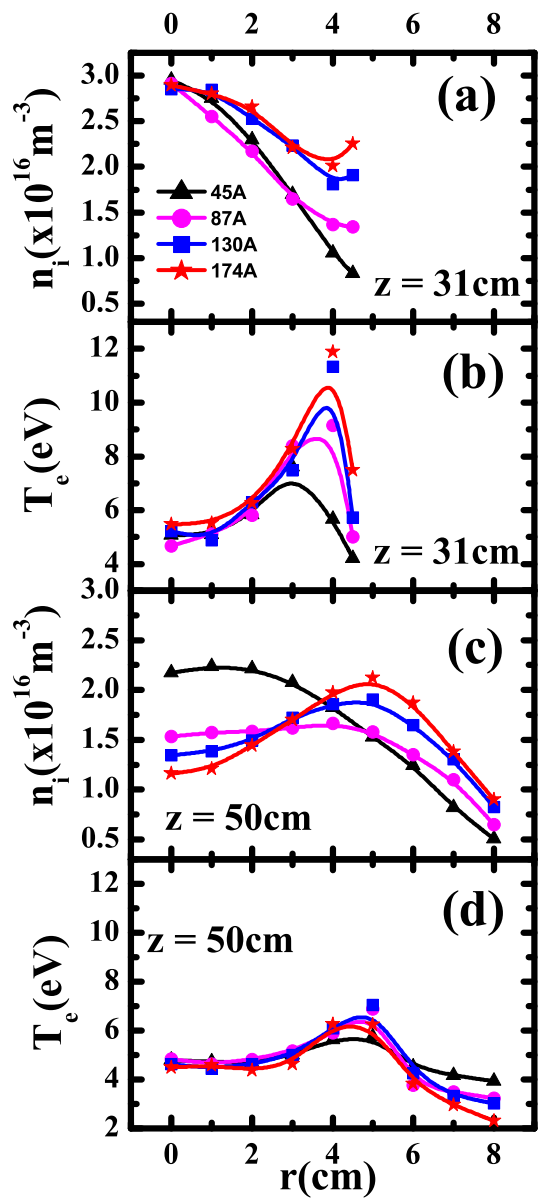}
\caption{Plasma density (a,c) and electron temperature (b,d) at locations of z = 31cm and z = 50cm, respectively, 
at rf power 200W and $1\times10^{-3}$ mbar argon fill pressure. For $I_B$, 45A (solid triangles), 
87A (solid circle), 130A (solid square) and 174A (solid star). Data are spline fitted for representation.} \label{fig:6}
\end{figure}

To observe the evolution of the radial profiles, these are obtained with different applied magnetic field 
strengths and are shown in Fig. \ref{fig:6}. Four values of the electromagnetic current $I_B$ are chosen carefully: 
$I_B$ = 45A,87A,130A and 174A for this purpose. 
Figs. \ref{fig:6}a and b show the profiles of electron density and temperature at
$z_{before}$ whereas Figs. \ref{fig:6}c and d show those at $z_{after}$
at fixed rf power of 200W and argon pressure $1\times 10^{-3}$ mbar. 
From Fig. \ref{fig:6}a it is seen that the radial profiles of plasma density in the source chamber at $z_{before}$
have a peak on axis for all values of $I_B$ and the ratio of the center and edge ($r\sim4$ cm) densities 
decreases as $I_B$ is increased. Here the effective electron temperature is nearly same at the 
axis for all values of $I_B$ and gradually increases towards the edge. The effective electron temperature 
peaks at an outer radial location and the peak value increases with IB (Fig. \ref{fig:6}b).
Increase in the plasma density at the outer edge follows the increase in the effective electron 
temperature there.  It is seen from Figs. \ref{fig:6}a and b that radial profiles of plasma parameters in 
the source chamber are not substantially different above $I_B$ value of 87A.
On the other hand, in expansion chamber at $z_{after}$, though the plasma density remains peaked at 
the center for $I_B$ = 45A, but it becomes hollow for $I_B$ values greater than 80 A (Fig. \ref{fig:6}c).
The density peak occurs off axis at $r\sim~$5 cm. 
At this position ($z_{after} = 50$ cm) the magnetic field is diverging (Fig \ref{fig:2}). 
The percent of hollowness, that is, the ratio of center-to-peak density is also increases with $I_B$. 
The nature of electron temperature profiles with $I_B$ follows the same 
behavior as in source chamber, that is, the temperature peaks at outer radial location, (Fig. \ref{fig:6}d). 
The ratio of peak-to-center electron temperature increases with $I_B$. 
The peak location more or less coincides with that of density here at z = 50 cm.
Value of electron temperature remains nearly same about (4.5 $\pm$ 1 eV) on axis at the locations
before and after the magnetic expansion, however,
it becomes significantly different at outer radial locations. 
It is seems that the electrons having temperature of 9-12 eV at outermost radial location ($r\sim$4 cm)
in the source chamber are not available in expansion chamber for all values of $I_B$. 
The electrons in expansion chamber at z = 50 cm, have maximum
temperature of 6 - 8 eV at $r\sim$5 cm which corresponds to the electron temperature at $r$ = 3 - 3.5 cm 
in source chamber
(z = 31 cm). 
\begin{figure}
\centering
\includegraphics[width= 7cm]{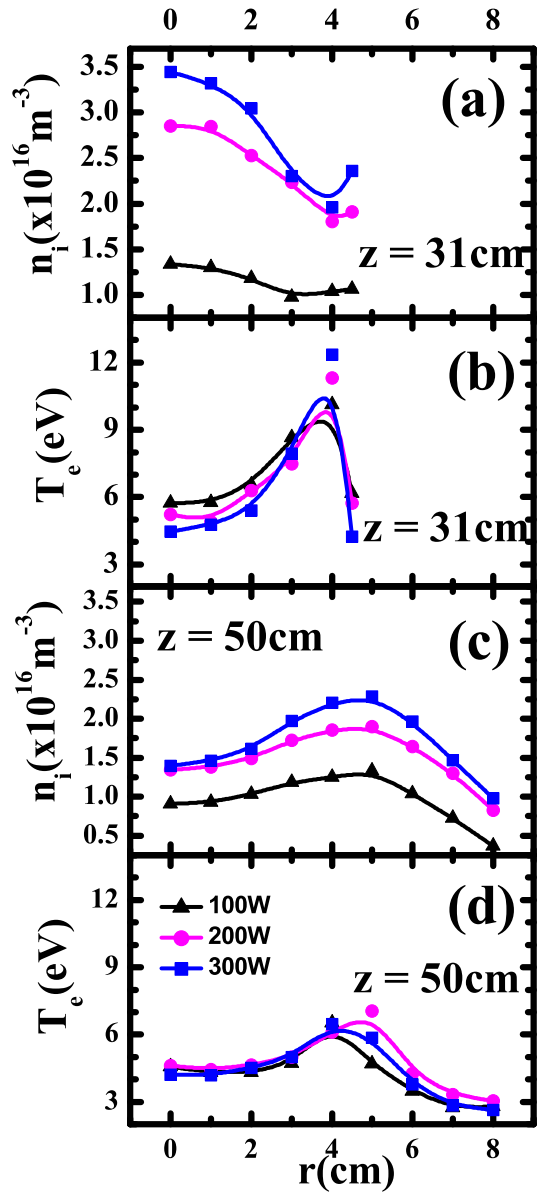}
\caption{Plasma density (a,c) and electron temperature (b,d) at locations of z = 31cm and z = 50cm, respectively, 
for fixed $I_B$ = 130A and argon fill pressure $1\times10^{-3}$mbar, for different 
rf power, 100W (solid triangles), 200W (solid circle) and 300W (solid square).
Data are spline fitted for representation.} \label{fig:7}
\end{figure}

\begin{figure}
\centering
\includegraphics[width= 7cm]{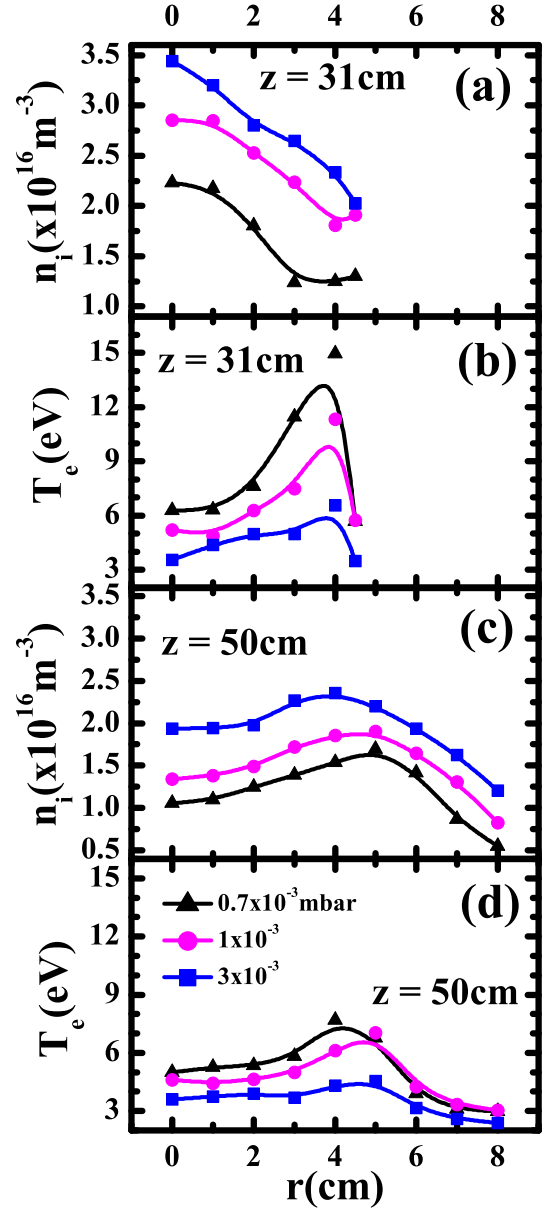}
\caption{Plasma density (a,c) and electron temperature (b,d) at locations of z = 31cm and z = 50cm, respectively, 
for rf Power 200W and $I_B$ = 130A, for different pressures,  
$0.7\times10^{-3}$ mbar (solid triangles), $1\times10^{-3}$ mbar (solid circle) and
$3.3\times10^{-3}$ mbar (solid square). Data are spline fitted for representation. } \label{fig:8}
\end{figure}

Establishing the fact that the electron density profile is center-peaked in the expansion chamber at low magnetic
there, but becomes hollow above a minimum magnetic field, we try to explore whether this hollow nature depends
on the other relevant parameters of the plasma source, like rf power and fill pressure. 
Fig. \ref{fig:7} and Fig. \ref{fig:8} show the results of such investigations.
First, we fixed the magnetic coil current $I_B$ and filling pressure of argon at 130A and $1\times10^{-3}$ mbar,
respectively and vary the operating rf power. 
The radial variations of plasma density and electron temperature are shown in Fig. \ref{fig:7}. 
In the source chamber the radial profiles of plasma density remains 
center peaked as before for all rf power values; however, the density increases throughout the profile with rf power,
(Fig. \ref{fig:7}a).
The effective electron temperature peaked at radially outward location (at around 4 cm) and does not
vary substantially with rf power, (Fig. \ref{fig:7}b).
In expansion chamber the plasma density shows hollow profile for all rf powers (Fig. \ref{fig:7}c),
though overall density density increases to some extent with rf power. 
Electron temperature is also peaked radially outward 
at r $\sim$ 5 cm (Fig. \ref{fig:7}d) and here also does not vary much with rf power.

Next, the magnetic coil current $I_B$ and rf power are fixed at 130A and 200W, respectively, and the 
filling pressure of argon is varied. The resulting radial plasma density and electron temperature profiles
are depicted in Fig. \ref{fig:8}. 
In the source chamber (z = 31 cm) the plasma density remains center peaked as before for all pressures
(Fig. \ref{fig:8}a) and the density increases throughout the radius with increasing filling pressure as expected.
The electron temperature profiles (Fig. \ref{fig:8}b) also remained peaked at radially outer location but
the temperature is reduced throughout the radius with increasing pressure.
In expansion chamber the hollowness in the density profile remains and the density increases over the whole 
profile with filling pressure, but with increasing pressure the hollowness tends to be suppressed, 
(Fig. \ref{fig:8}c). 
Similar behavior is observed with electron temperature (Fig. \ref{fig:8}d) which remains peaked off-axis,
but decreases with increasing pressure at all radii.

\section{\label {sec:level} Discussion:}

The experimental results presented in Fig. \ref{fig:6} to Fig. \ref{fig:8} can be summarized into two 
broad aspects: A) irrespective of the values of the electromagnet coil current, the radial profiles 
of the electron temperature remain peaked off-axis in both the source and expansion chambers, though 
the values are different, and B) the electron density is center-peaked in the source chamber irrespective
of the coil current, however, in the expansion chamber it started to become hollow above a critical value 
of the coil current. In the following we elaborate on these findings and explore the underlying 
physics behind these observations.
\subsection{\label {sec:level}  Off-axis electron heating:}
The region of radially outward electron temperature peaking  
solely depends on operating conditions of helicon plasma discharge. 
Depending on input  
operating parameters (rf power, gas pressure and applied magnetic field), helicon discharge 
can be in capacitive (E), inductive (H) or Helicon wave (W) mode \cite{ref_10,ref_11,ref_12,ref_13}. 
The radial structures of plasma density and electron temperature are also dictated 
by the dominant mode of discharge operation. 
In our operating conditions the discharge is predominantly in inductive mode.
The off-axis peaking of the electron temperature in source chamber at z = 31 cm (Fig. \ref{fig:6}-\ref{fig:8})
indicates that a local electron heating mechanism must exist near radial plasma boundary.

In an insulating source discharge tube where the helicon antenna is placed around, a skin layer is formed near the
radial boundary where most of the rf- power is transferred to electrons \cite{ref_15,ref_16} and plasma
is produced by electron impact ionization. 
The collisionless skin depth ($\delta = \frac{c}{\omega_p}$) with a plasma density of
$3\times10^{16}m^{-3}$ is about $\sim$2.5 cm, which is nearly
half the radius of our source chamber.
As there is an external magnetic field, so the bulk electrons are unable to freely diffuse from the skin layer 
or in other words the electrons which are present in skin layer are restricted by cross-field diffusion mechanism.    
This is the reason why we see the temperature profiles of electrons in the source region (shown in
Fig. \ref{fig:6} - \ref{fig:8}) to be peaked off-axis.
The increase of off-axis electron temperature by external magnetic field 
(Fig. \ref{fig:6}b), is due to confinement of these electrons in the skin layer by
their small gyroradii, which are 
about 0.44 mm for an average energy of 12 eV in 300G (z = 31 cm).
It is generally seen that some of the electrons in this region, instead of losing energy by ionization,
acquire high energy and form a tail in electron energy Maxwell distribution in the source chamber near the 
antenna location.
These high energy electrons are transported by the 
last diverging peripheral magnetic field lines emerging from open exit of source into the expansion chamber 
as shown in Fig. \ref{fig:2}a.
The electrons that are very close to the source radial boundary at r $\sim4$ cm are not 
transported into the expansion, since these electrons are tied to magnetic field lines 
and follow the curvature of magnetic field (Fig. \ref{fig:2}a) to hit the source peripheral wall
near to open exit and lost. 
Hence the electron energy distribution including the high energy tail (Fig. \ref{fig:4})  
in the expansion chamber at r $\sim$5 cm, 
corresponds to that at the radial location of r $\sim$ 3 - 3.5 cm in source chamber 
(shown by circle in Fig. \ref{fig:2}a). 
The external magnetic field plays an important role for maintaining of the electron temperature gradient in our 
experiment as inhibition of
cross field diffusion restricts interaction among electrons of all regions.
\begin{figure}
\centering
\includegraphics[width= 7cm]{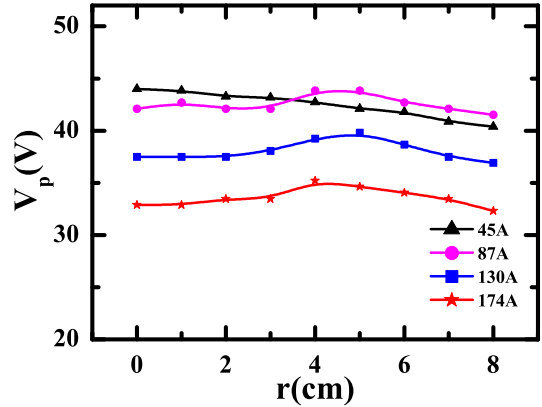}
\caption{Radial plasma potential at z = 50cm in expansion chamber at 200W 
and $1\times10^{-3}$mbar argon pressure. 45A solid triangles, 
87A solid circle, 130A solid square and 174A solid star. Data are spline fitted for representation.} \label{fig:12}
\end{figure}
\subsection{\label {sec:level}  Hollow density formation:}

Formation of hollow density profile in expansion chamber can either be due to radially outward
transport of plasma or due to additional ionization off-axis. Fig. \ref{fig:12}
shows the local plasma potential profiles in the same 
conditions as in Fig. \ref{fig:6}c and d. The radial electric field in
expansion chamber at z = 50 cm is very weak,  $ < 0.7 V/cm$ and it decreases with increase in external 
magnetic field. 
So, the possibility of outward radial transport of plasma due to 
$\bold E\times \bold B$ drift can be ruled out in our experiment. 
In reference 18, we established the presence of additional
off-axis ionization in expansion chamber.
It was shown there that poloidal rotation of electrons due to the gradient-B drift plays a crucial role in the 
off-axis ionization. The mechanism is briefly stated here.
The temperature of high energy tail electrons in expansion chamber ranges from 12-15 eV and their
population generally \cite{ref_probe_ana1,ref_probe_ana2} is about 15 - 20 $\%$ of the total electron density. 
So, there are a substantial number of electrons having energy of 20 - 50 eV present off-axis. 
However, as their ionizing collision length at argon pressure of $1\times10^{-3}$ mbar 
is $\sim150$cm, which is much larger than the system size,
so they cannot produce additional plasma off-axis moving along the axis.
On the other hand, the hollowness in density profile is found only after the magnetic divergence \cite{ref_19}, 
where a strong gradient in external magnetic field ($\triangledown B$) is present.  
It has been shown there that these high energetic electrons undergo a high rotational motion by
gradient-B drift in azimuthal direction. The motion is localized near the magnetic field divergence
and provides sufficient time to cause the ionizing collisions with the neutrals leading 
to off-axis additional ionization within few rotations.
\begin{figure}
\centering
\includegraphics[width=7cm]{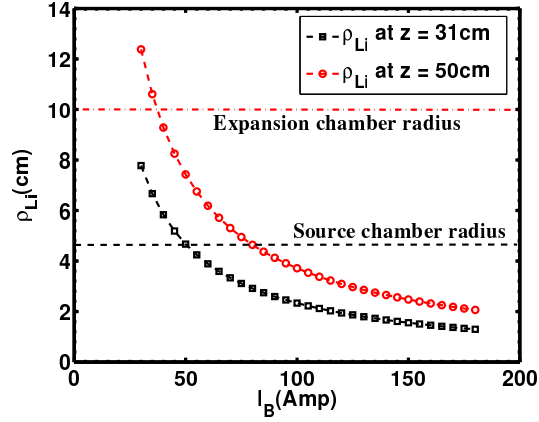}
\caption{ Ion larmor radius at z = 31cm source (open square) and z = 50 cm expansion (open circle) chamber;
for different values of coil currents $I_B$. The magnetic field strength has $\sim$1.75 G/A and $\sim$1.1 G/A
at z = 31 cm and 50 cm respectively. Two horizontal dotted dash and dash lines correspond to
expansion and source chamber radii respectively.} \label{fig:5}
\end{figure}

However, in the present experiment the interesting result that is obtained is a critical value of 
the magnetic field above which the hollowness in plasma density in the expansion chamber
is found (Fig. \ref{fig:6}c). 
The radial profile of plasma density in the expansion  
chamber is center peaked for magnetic field of 50G, (corresponding to coil magnetizing current $I_B$ = 45A). 
So it seems the grad-B drift effect is an essential condition, but not sufficient tone o form radial 
hollow density profile. 
In order to investigate the existence of the critical magnetic field, we plot the calculated ion Larmor
radii at the two axial locations (where the experimental results of Figs. \ref{fig:6}-\ref{fig:8} are obtained) for
different coil currents $I_B$ in Fig. \ref{fig:5}. 
The values of the source and expansion chamber radii are also indicated in the figure for reference.
Interestingly, it should noted here that out of the four values of $I_B$ (45A, 87A, 130A and 174A) for 
which the profiles of Figs. \ref{fig:6}-\ref{fig:8} are obtained, only for $I_B$ = 45A the ion Larmor radius in the
expansion chamber is close to the chamber radius, that is, the ions are here not magnetized; for all other coil
currents ions are magnetized. 
The change in electron density profile depends on the confinement of ions through ion gyroradius 
in expansion chamber 
radius of 10 cm. For magnetic field of 50G at z = 50 cm, as the ions gyro radius $\sim$ 9 cm, that is,
close to the system radius.
So though the high energy electrons are confined and the gradient-B effect produces off-axis ionization,
but ions are not magnetized; as a results extra off-axis plasma is lost to the wall by ambipolar field. 
The density remains center-peaked due to quasi neutrality and non-magnetized ions. For 95G of magnetic field 
(corresponding to 87A) the ion gyroradius becomes $\sim$ 4.8 cm; here ions are magnetized and density 
is flattened on that scale. While for still higher values of magnetic field the ion gyroradius become much 
smaller than the system dimension and electron density starts to become hollow.

\begin{figure}
\centering
\includegraphics[width=7cm]{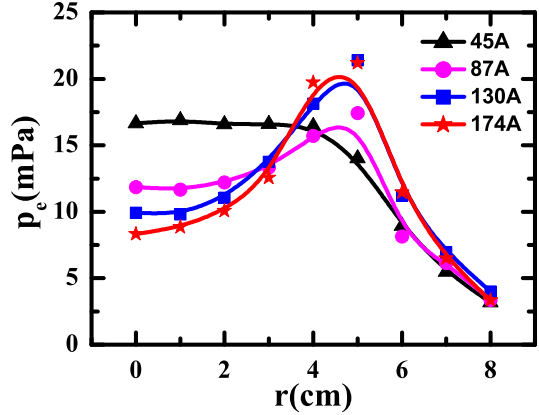}
\caption{Radial electron pressure $p_e$ profile, at z = 50cm in 
expansion chamber at 200W rf power and $1\times10^{-3}$mbar argon fill pressure. 45A (solid triangles), 
87A (solid circle), 130A (solid square) and 174A (solid star). Data are spline fitted for representation. } \label{fig:13}
\end{figure}

The hollowness in plasma density and off-axis peaking of electron temperature leads 
to hollow electron pressure ($p_e = n_0k_BT_e$) profile. The calculated
radial profile of electron pressure from the measured profile of plasma density and electron temperature is 
shown in Fig. \ref{fig:13}. 
The electron diamagnetic current($j_{De} = \frac{1}{B_z}\frac{\partial p_e}{\partial r}$)
flows in anti-diamagnetic direction at around $r\sim5$cm due to 
hollow pressure profiles ($I_B$ = 130 and 174A), while $j_{De}$ at $r>5$cm is in diamagnetic direction.
This analogy is consistent with recent previous measurements \cite{ref_azimu_Jd_1}.
Formation of hollow structure within magnetically expanding plasma 
causes the reduction of total thrust 
due to anti-diamagnetic direction of azimuthal current, 
since it leads to revers role of Lorentz force in presence of radial component of applied diverging
magnetic field\cite{ref_7,ref_Jd_direction}. 

 
\section{\label {sec:level}  summary and Conclusion:}

In our helicon plasma source based linear device with expanding plasma geometry and diverging magnetic field, 
the radial density profile is found to be peaked on-axis and evolves into a hollow profile in the expansion
plasma as the diverging magnetic field at the axial location is increased beyond a critical value. 
It is seen that the hollow density profile is formed only when both electrons and ions are magnetized at
this location and where the presence of magnetic divergence helps to increase the confinement of 
hot electrons due to the azimuthal rotation caused by the grad-B drift. On the other hand, 
irrespective of the plasma operating conditions of the experiment, the source radial density 
profile is center-peaked. The effective electron temperature is peaked radially outward in 
the source region for all values of the magnetic field due to the rf skin heating effect near the helicon 
antenna, where an hot electron component is also generated. The hot electrons and the temperature 
profile hollowness are transported to the expansion chamber along the divergent magnetic field 
lines and the hollowness becomes more prominent for higher magnetic field.

\nocite{*}
\bibliography{annula_plasma.bib}

\end {document}